# Energy frontier lepton-hadron colliders, vector-like quarks/leptons, preons and so on

Saleh Sultansoy

TOBB University of Economics and Technology, Ankara, Turkey

ANAS Institute of Physics, Baku, Azerbaijan

ATLAS, LHeC and FCC Collaborations

Member of Plenary ECFA

**Abstract**

First of all, an importance of the LHC/FCC based energy frontier lepton-hadron and photon-hadron colliders is emphasised. Then arguments favoring existence of new heavy isosinglet down-type quarks and vector-like isosinglet or isodoublet leptons are presented, following by historical arguments favoring new (preonic) level of matter. The importance of Super-Charm factory and GeV energy proton linac for Turkey's national road map is argued. Finally, several recommendations for ESPP2020 are suggested.

**Input to the European Particle Physics Strategy Update**

**Contents:**

İntroduction

1. LHC/FCC based energy frontier lepton-hadron and photon-hadron colliders

2. Flavour Democracy calls for vector-like leptons and quarks

3. History calls for Preons

4. Some words on Turkey's national road map

Conclusion and recommendations

References

## Introduction

This paper reflects current status of our opinions on European Strategy for Particle Physics (for earlier contributions see [1, 2] and references therein).

In following section arguments favoring LHC/FCC based energy frontier lepton-hadron and photon-hadron colliders are presented. Section 2 is devoted to present status of Flavour Democracy Hypothesis. Historical arguments in favors of new level(s) of matter substructure are given in Section 3. Section 4 is devoted to present status of linac-ring type Super-Charm Factory, which is the cornerstone of TAC (Turkic Accelerator Complex) project. Finally, several recommendations are listed.

## 1. LHC/FCC based energy frontier lepton-hadron and photon-hadron colliders

Parameters of Proposed Future Colliders at CERN are summarised in [3] (INSTRUCTIONS FOR THE PREPARATION OF CONTRIBUTIONS TO CERN REPORTS). Below several comments on lepton-hadron options are presented.

Certainly, construction of QCD Explorer stage of LHeC is mandatory [4] in order to provide precision pdf's for HL(HE)-LHC, FCC-hh and/or SppC, as well as to clarify QCD basics (especially small $x_g$ region related with confinement). It should be noted that L of order of $10^{32}$cm$^{-1}$s$^{-1}$ is quite enough for these purposes (high x region will be explored by EIC, USA). Concerning energy frontier stage of LHeC, its necessity will depend on the results from the HL-LHC.

In [3] ERL60 is considered as sole option for both LHC and FCC based lepton-hadron colliders (see also [5]). It is reasonable to consider ERL60 as baseline (not sole) option for LHeC. Nevertheless, other opportunities [6-8], especially µp and µA colliders, should be studied in details as well. As for FCC, ERL60 can't even be the baseline choice: beam energy asymmetry is 833! Different options for FCC based energy frontier ep, µp, eA, µA, γp and γA colliders have been proposed in [9] (see also [23, 24]), where ILC and PWFA-LC have been considered for electron (0.5 TeV and 5 TeV, respectively) and photon beams, µC with $E_\mu$ values 0.75 TeV and 1.5 TeV has been considered for muon beam. With moderate upgrade of the FCC parameters and IP design, L of order of $10^{33}$cm$^{-1}$s$^{-1}$ seems to be achievable for all options.

Let me emphasize that even with L~$10^{32}$cm$^{-1}$s$^{-1}$ FCC based energy frontier lepton-hadron and photon-hadron colliders search potential for BSM physics is far beyond corresponding lepton colliders and exceed the potential of FCC itself for a lot of phenomena (see [10-12] and references therein for LHC, ILC and ILC-LHC comparison. Similar work should be performed for FCC!)

## 2. Flavour Democracy calls for vector-like leptons and quarks

Mass and mixing patterns of the SM fermions are among the most important issues, which should be clarified in particle physics. In recent interview published in CERN Courier [13], Steven Weinberg emphasized this point: "Asked what single mystery, if he could choose, he would like to see solved in his lifetime, Weinberg doesn't have to think for long: he wants to be able to explain the observed pattern of quark and lepton masses". In our opinion, Flavor Democracy (see review [14] and references therein) could provide an important key to solve this mystery.

Within the framework of the SM, Flavour Democracy predicted the existence of the fourth family. Today, SM4 (with one Higgs doublet) is excluded by Higgs boson properties, but this is not the case for general chiral fourth family (see [15] and references therein). On the other hand, addition of heavy isosinglet down-type quark and isosinglet (or vector-like isodoublet as predicted by $E_6$ GUT) lepton to SM3 give opportunity to obtain masses of charged leptons and quarks of the 2nd and 3rd family due to small deviations from full Flavor Democracy [16].

### 3. History calls for Preons

Standard Model (SM) has proven its reliability by the experimental verifications of its particle content in the recent decades. SM puzzle has been completed by the discovery of Higgs boson. However, SM seems not to be the end of the whole story. There are still many unsolved problems that are out of the scope of the SM and especially the large number of currently known elementary particles becomes more of an issue. For this reason a lot of BSM models have been proposed including extension of scalar and fermionic sectors of SM, enlargement of SM gauge symmetry group, SUSY, compositeness (preons [17], see, also [18] and references therein), extra dimensions etc. Keeping in mind historical development of fundamental building blocks of matter, the search for preonic models seem to be quite natural. This development is summarized in Table I from [19].

Table 1. Historical development of "fundamentality".

| Stages | 1870-1930s | 1950-1970s | 1970-2030s |
|---|---|---|---|
| Fundamental Constituent Inflation | Chemical Elements | Hadrons | Quarks, Leptons |
| Systematics | Periodic Table | Eight-fold way | Family Replication |
| Confirmed Predictions | New Elements | New Hadrons | $l^*$, $q^*$, $l_8$, $q_6$, …? |
| Clarifying Experiments | Rutherford | SLAC-DIS | LHC or rather FCC/SppC? |
| Building Blocks | p, n, e | Quarks | Preons? |
| Energy Scale | MeV | GeV | (multi-)TeV? |
| Impact on Technology | Exceptional | Indirect | Exceptional? |

As was mentioned in [20]: "In my opinion, enormous efforts devoted to SUSY related topics divert our attention from possible alternatives." Today, decades of SUSY-cracy is about to come to an end. Indeed, the were two reasons (in addition to its beauty) for SUSY: unification of coupling constants and protection of the Higgs boson mass. First argument works only for SU(5) GUT, the second one falls off the agenda with the absence of squarks and gluinos with masses below TeV. Therefore, that is the time to turn our eyes to the Preonic models!

Potential of exploration of preonic models at the FCC based pp, ep, $\mu$p and $\gamma$p colliders was reviewed in [21] (see also presentations at workshop titled "New Building Blocks of Matter: Preons" [22]).

## 4. Some words on Turkey's national road map

25 years ago Linac-Ring Type c-$\tau$-Factory was proposed [25] as regional project for particle physics (region means Mid East + Balkans + Caucasus + Central Asia). TAC (Turkic Accelerator Complex) project was supported by DPT (Turkish Planning Committee, now Ministry of Development) in 1997-2015 [26]. Originally project included also Synchrotron Radiation Source based on positron ring, later FEL based on GeV electron linac and GeV energy proton linac were added during the feasibility study stage (1997-2000). Actually, Super-Charm factory and proton linac are most importanat parts of the TAC.

In addition, FEL based on 20-40 MeV energy normal conducting e-linac was proposed as an education facility for TAC in 2005. With the wrong decision taken in 2007, super-conductor technology was chosen for e-linac and the training facility was transformed into a user facility (so called TARLA project [27]). Unfortunately, the TARLA project blocks the development of accelerator technologies in Turkey.

TAC Super-Charm factory related work was de-facto stopped in 2003 (for details see [24, 28]). These studies should be restarted. Particle factory and GeV energy proton linac must form the main axes of the national road map. This road map should also include full membership to CERN, as well as real partnership with DESY, ESS, MYRRHA and so on.

**Conclusion and recommendations**

In light of the above arguments, I make the following suggestions:

i) Organization of special working group on the LHC/FCC based energy frontier lepton-hadron and photon-hadron colliders (maximally disentagled from ERL60/PERLE governance) will be very useful,

ii) ATLAS and CMS should pay special attention to new heavy isosinglet down-type quarks and vector-like (isosinglet or isodoublet) leptons,

iii) In the LHC and FCC physics research programs, preonic models should be covered at least as much as SUSY,

iv) TAC Super-Charm Factory needs support from HEP community, as well as governing bodies (ECFA, ICFA and so on).